# Protein-polysaccharide interactions and aggregates in food formulations

Luigi Gentile

*Department of Chemistry, University of Bari "Aldo Moro", Via Orabona 4, Bari I-70126, Italy*
*Center of Colloid and Surface Science (CSGI) Bari Unit, Via Orabona 4, Bari I-70126, Italy*

**Abstract**

The protein-polysaccharide combinations that lead to electrostatic complex and coacervates formation are the object of extensive research using both layer-by-layer and mixed emulsion approaches. The protein-polysaccharide conjugates demonstrated interesting physico-chemical properties as stabilizers and emulsifiers as well as texture modifiers in food products. Furthermore, they are potential optimal nutrient delivery systems. Their complex behavior due to several factors such as pH, ionic strength, concentration, heat, and mechanical treatments is the main reason behind the continuous growth of the research field. The review is reporting same recent advances on the topic, along with an overview on the possible interactions between protein and polysaccharide, from Maillard reaction to enzymatic crosslinking passing through coacervates.

"Keywords: protein-polysaccharide, coacervates, Maillard, layer-by-layer, mixed emulsions."

## 1. Introduction

The organization of food constituents at multiple spatial scales and their interactions is the so-called food structure [1]. The food structure is a function of the food ingredients and production process. The composition of the food and its structure determines the texture, the perceived attribute, and the mechanical properties [2]. A food go through several steps during consumption. The initial mechanical breakdown of the food structure starts with the mastication and mixing using the tongue. The second step is the lubrication, hydration, and dissolution due to the enzymatic action of the saliva, the small broken piece of food is then converted to bolus and swallowed through the esophagus into the stomach. Food industries are facing the problem to make foods healthier and at the same time not diminish sensory quality [3,4]. Moreover, nutrient delivery systems, microencapsulation and protection of active ingredients and as a consequence dispersant agents are to be taken into account during structural design [5*]. Manufactured foods commonly exist in the colloidal state as emulsions, foams, gels, and dispersions. The food colloidal science investigates the influence of ingredient composition and formulation conditions on structure, stability, and mechanical properties [6*]. The texture concern is mainly the flow behavior for dispersions, while breaking force, breaking strain, and the size distribution of the newly formed particles are the main physical parameters that can be correlated to hardness, brittleness, crispness, crunchiness, and crumbliness for solid-like foods. The mouthfeel of food emulsions is strongly influenced by the type, concentration and interactions of the particles and macromolecules present. The perceived fattiness, creaminess and thickness of oil-in-water emulsions have been found to increase as the droplet concertation increases [7]. The creaminess was also found to depend on droplet size and to the emulsifier type might due to droplet flocculation and emulsion viscosity. During the consumption of some food emulsions, there is a cooling sensation associated with melting of emulsified fat in the mouth due to the endothermic enthalpy change associated with fat crystal melting [8,9].

To obtain proper texture and to stabilize the food products several emulsifiers, solubilizes and dispersing agents are adopted. The focus of this review will be the protein-polysaccharide combinations in conjugates or complex structures and their multiple applications in food industries. The protein-polysaccharide combinations are particularly interesting since their ability to change product shelf life by varying food texture, i.e. rheological properties of food colloids [10,11,12**], for such reason they have been the object of intense research [13,14**,15–19]. There are several interesting reviews on the matter [2,5*,6*,12,15**,19,20*] along with a vast literature. This review should not be considered fully comprehensive, while the intent is to provide a general overview of the topic along with the most recent findings.

Food proteins can avoid flocculation of emulsion droplets since they have a strong tendency to adsorb onto hydrophobic-hydrophilic interfaces and subsequently unfold (or partially unfold) forming relatively thin adsorbed surface layers (~2–6 nm) that generates electrostatic and steric stabilization [21]. The van der Waals attraction force between colloidal particles overcome the electrostatic repulsion, at the same time the protein charge is the main reason for which colloidal stability can be lost. At the isoelectric point (pI) the negative and positive charges are balanced, reducing repulsive electrostatic forces, and causing aggregation and precipitation [22,23]. Moreover, several studies on proteins demonstrated how pH, ionic strength, concentration, and heat treatment influence the pI [24,25].

The protein and polysaccharide combinations allow designing an amphiphilic conjugate to be strongly anchored to the oil–water interface via the protein's hydrophobic regions, leading to a viscoelastic layer, with the non-adsorbing polysaccharide (copolymer) region to provide enhanced steric stabilization [14**] that can lead to gelling behavior.

There are two main kinds of interactions between polysaccharides and proteins: covalent or non-covalent bonds. The covalent bond is obtained through a Maillard-type reaction that it is leading to protein-polysaccharide conjugates with elevated heat stability. However, the reaction conditions, such as pH and temperature, should be properly settled to obtain the desired reaction. The next paragraph will address the reaction condition needed to a successful reaction. The driving forces for the non-covalent bonds are electrostatic, hydrophobic, H-bonding and Van der Waals interactions, such forces can generate coacervates that are useful tools to change food texture and encapsulate active compounds. One relevant topic of this review is the combination of the Maillard reaction with the electrostatically driven aggregates such as coacervates.

## 2. Maillard reaction: covalent bonds

A limited number of polysaccharide attaches to protein due to the steric hindrance of the macromolecule, usually only a couple of molecules attach to folded proteins such as ovalbumin and lysozyme, while several polysaccharides attach to unfolded proteins like casein [26]. The key factor for such reaction is related mainly to the presence of lysine that undergoes a Maillard type reaction with the reducing sugar of polysaccharides, under suitably low water activity and heat treatment over an incubation period of at least few hours [27]. One considerable limitation consists of deteriorates protein functionality if most of the lysyl residues are masked by the saccharide, as was observed in the conjugates of proteins with monosaccharides and oligosaccharides. However, the Maillard-protein-polysaccharide conjugates showed excellent emulsifying properties respect to conventional commercial emulsifiers, antimicrobial activity, and heat stability. Therefore, the conjugates are adopted for industrial applications [28]. It is relevant to report some safety issues highlight in a critical review by de Oliveira et al. [20*], where the essential point is that the Maillard reaction must be conducted properly to avoid the generation of harmful compounds [28,29]. The conjugates formed via the Maillard reaction consist of various glycoforms and are produced in combination with a large number of secondary products directly related to the reaction conditions (pH, temperature, humidity, etc…) [29,30].

Li and Etzel [31**] studying whey protein isolate and dextran underline that glycates dissociate by hydrolysis, returning to free un-glycated protein and polysaccharide due to the reversibility of the Schiff base linkage [32]. They determinate rate constants and equilibrium constants for glycate hydrolysis, temperature was obviously increasing the hydrolysis speed, as expected. The common processing temperature in thermal processing of foods is around 85 °C and such value could cause significant glycate hydrolysis. The authors suggested a food-grade reducing agent such as sodium borohydride and ascorbic acid after the formation of the glycated. Their approach consists of the conversion of the Schiff base into a stable secondary amine. Moreover, if the Schiff base reduction is done during glycate formation by the wet-heating method, the reaction will be driven towards the glycate. Kutzli et al. [33] investigated the glycation of pea protein isolate with maltodextrin through the first stage of the Maillard reaction after the physical structuring of the reactants by needleless electrospinning. Their results indicate that glycation of the pea protein with maltodextrin in electrospun fibers is able to improve techno-functional properties. However, they reported an increase in the browning index of the fibers with increasing heating time and temperature due to the formation of a Schiff base that can decompose after acidic hydrolysis to Amadori products. Their findings are supporting the current opinion related to the pI shift; they reported a shift from pH 4.05 ± 0.13 to pH 3.02 ± 0.16 for fibers heated at 65 °C/24 h and at 70 °C/24 h, respectively. The heating process was leading to conjugates with higher solubility compared to the unheated fibers over the pH range from 2 to 7.

Zha et al. [34*] focused their work on the properties and functionalities of conjugates formed between pea protein isolate and gum Arabic respect to the incubation time (0, 1, 3, and 5 day). The corn oil-in-water emulsions stabilized by the Maillard-protein-polysaccharide conjugates with 1 day incubation had greater stability against environmental stresses than those prepared by protein or protein-gum Arabic mixture. At longer incubation time (5 days) the extensive reaction can reduce the emulsification property of conjugates. The conjugates, formed through the controlled Maillard reaction, were also able to inhibit the formation of volatile compounds and prevent emulsion oxidation. The formation of the Amadori compounds monitored by UV-Vis absorbance at 304 and 420 nm was observed already during 1 day incubation, while the browning was observed to increase with the incubation time. Zha et al. demonstrated that controlled Maillard reaction with hydrophilic polysaccharides can modulate the solubility and functionality of poor water soluble plant proteins. On the contrary in Zhong et al. [35*] an incubation time of 5 days was chosen at pH 8, to form more conjugates [28] even though that implies the formation of Amadori compounds in higher extent [36]. However, they reported a relatively low change in the browning index; in fact, after 5 days it was the double respect to the first day. Zhong et al. focused on a conjugate made with oat protein isolate and Pleurotus ostreatus β-glucan via Maillard reaction under controlled dry-heating conditions. The dry conditions were most likely affecting to some extent the formation of Amadori compounds. In their analysis of the amino acid composition, cysteine and lysine were identified as the dominant Maillard reaction sites. After covalent binding with P. ostreatus β-glucan, the incompact surface structure and decreased surface hydrophobicity of oat protein isolate caused its increased solubility and emulsibility. The introduction of P. ostreatus β-glucan enhanced the thermal stability of oat protein isolate due to its extended secondary structure induced by the Maillard reaction. The macromolecular P. ostreatus β-glucan in conjugate forms long-range steric repulsion between the surfaces of emulsion droplets. In addition, it promotes the formation of a stable membrane around the oil droplets, which was helpful to increase the emulsifying activity of oat protein isolate [37–39]. Under the controlled condition, Maillard reaction was an effective way to improve the application potentials of oat protein isolate in food processing. At the structural level that the secondary structure of conjugates was altered by decreasing the contents of α-helix and β-sheet and increasing the contents of β-turn and random coil. The surface structure of conjugates was loose and porous.

## 3. Non-covalent bounds

Proteins can be positively or negatively charged, depending on the pH. Carboxylate polysaccharides get negatively charged at a pH range higher than its pKa. These electrical charges on the backbone of protein or polysaccharide chains are responsible for electrostatic interactions [40,41]. Moreover, hydrogen bonding and hydrophobic interaction play also a role in the stability of the protein-polysaccharide aggregates [42].

The protein-polysaccharide can exist in a single-phase system, i.e. the *associative phase separation*, for interacting biopolymers, leads to soluble complex (coacervation) or the *segregative phase separation*, for non-interacting biopolymers, leads to uniformly distributed components throughout the medium. They can also exist in insoluble precipitated for interacting biopolymers or the *segregative phase separation* can lead to distinct phases (2-phase system) [43**].

There are essentially two ways to adopt protein-polysaccharide combination to stabilize emulsions: the so-called layer-by-layer approach consists in the addition of charged polysaccharide to a primary protein stabilized emulsion; the other approach is sometimes termed as mixed emulsions that involve the addition of protein-polysaccharide complexes in aqueous solution and subsequently homogenization.

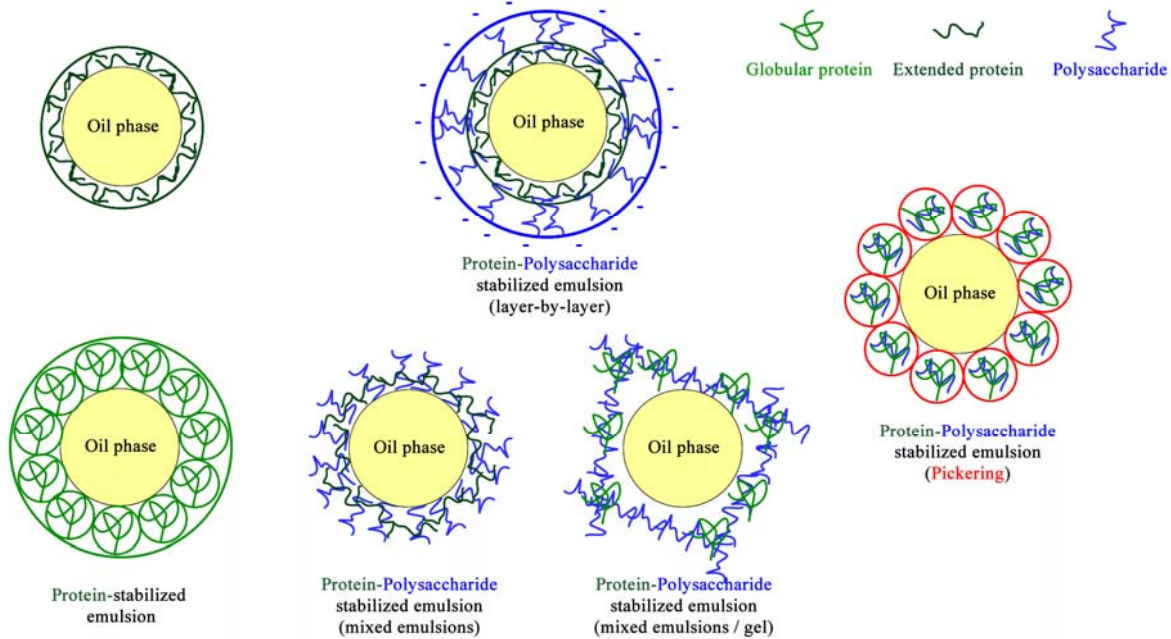

**Figure 1**. Schematic overview of protein and protein-polysaccharide stabilization of oil droplets in an oil/water emulsion. The mixed emulsions can lead to several kind of gel network (not represented). The extended protein could be in the random coil or partially folded state.

*3.1. Coacervates*

Food proteins with an isoelectric point around 5 can form complex coacervates with anionic polysaccharides such as pectin (pI around 3.5) in the intermediate-pH region, where the two macromolecules carry opposite net charges: pH above the pI of the polysaccharide but below that of the protein. Usually, simple coacervation is referred to a single biopolymer in the system, while when the interactions occur between two biopolymers the process is known as complex coacervation [44]. A common example of electrostatic complex formation in the food industry is that based on the interactions of globular positively charged proteins and oppositely charged ionic polysaccharides. It is worth mentioning that there are many examples in the literature where both polymeric polyions are proteins [45,46]. More information on protein–protein coacervates is reported in recent review articles [43**,47]. Interestingly, like-charged coacervates have been reported even between two positively charged polyelectrolytes by overcoming longer-range electrostatic repulsion [48]. In general, coacervates are reversible entities that form and segregate as the solution and/or environmental conditions are modified [49–51].

Lan et al. [52] investigated the complex coacervation between the pea protein isolate and the sugar beet pectin, a spontaneous exothermic process, mainly due to the electrostatic interaction and the hydrogen bonding between nonspecific amino groups of the pea protein isolate and carboxylic groups of the sugar beet pectin. They demonstrated that a state diagram could explicitly identify the three characteristic pH values ($pH_{\varphi 1}$, $pH_{opt}$, and $pH_{\varphi 2}$) of the complex coacervation. $pH_{\varphi 1}$ shifts from 3 to 5.5 along with structural changes as the pea protein isolate and sugar beet pectin mixing ratio increases from 1:1 to 20:1. When electro-neutrality is achieved, around the $pH_{opt}$, stronger and denser structure with greater storage modulus is formed. On the other hand, when the pH is near the $pH_{\varphi 2}$ the coacervates show smooth inner pore surfaces with homogeneous large pore size distribution. These kind of structural changes lead to different mechanical behavior, Chang et al. [53] investigated the rheological and microstructural properties of the canola protein isolate−chitosan coacervates at several mixing ratios and pHs. The elastic modulus, $G'$, was found to be higher than the viscous modulus, $G''$, for all the investigated mixing ratio and pHs, with the highest value of $G'$ at a mixing ratio 16:1 and pH 6.0 due to the high strength of electrostatic interaction and thick-walled, sponge-like, less porous microstructure. The canola protein isolate−chitosan complex coacervate phase formed at mixing ratio of 16:1 and pH 6.0 exhibited glassy consistency at low temperatures and rubbery consistency above its glass-transition temperature. For all the investigated systems, a predictable, shear-thinning behavior was also reported.

Wu et al. [54*] demonstrated that soybean protein isolate with the cationic polysaccharide chitosan or with carboxymethyl cellulose (anionic polysaccharide) can form stable coacervates when heating their mixtures at critical pHs. The complex particle formation was polysaccharide-type independent, while the particle size, polydispersity index, pH sensitivity, and ionic strength sensitivity were polysaccharide-type dependent. They also reported that when the soluble complex was heated at a specific temperature, it was not returning to the original state after cooling down.

The formation of stable coacervates is strictly related to the external condition such as pH, furthermore, they might not be stable to temperature changes.

Huang et al. [55*], on the same system, offered an interesting bridge between the Maillard reaction and the electrostatically driven coacervates. Their study was focusing on the occurrence of Maillard reaction during coacervation at 50 °C, 70 °C, or 90 °C for 12 h in the presence of maltose. In their finding as the extent of Maillard reaction increased, the $-NH_2$ groups of chitosan (initially positively charged) were consumed by the reaction with maltose increasing the negative charge. The coacervates exhibited decreased aggregation and higher stability due to the steric repulsion and only partially due to the repulsive interactions generated by the negative charge on the surface. The coacervates size was growing with temperature increasing up to ~5 γm at 90 °C, as well as the zeta-potential up to ~ -16 mV. However, coacervates have usually a higher absolute value of zeta potential [52,54]. Huang et al. did not take into consideration that 90 °C could cause significant glycate hydrolysis as reported by Li and Etzel [31**], even though they report indirect observations such as furosine and color tests, partially proving the final stage of the Maillard reaction. Huang et al. reported that the investigated coacervation temperatures improved the microencapsulation efficiency of Vitamin E and microencapsulation yield.

*3.2. Layer-by-layer*

An interesting way to promote polysaccharides adsorption at the interface is to use their possible electric charge to attract them to an already deposited layer of opposite charge on the surface. The idea owes its origins to the so-called layer-by-layer deposition process [56], to form multi-layers on macroscopic surfaces. In each stage of the process, the previous solution is washed. Then a new solution, containing polymers of opposite charge respect to the previous layer, is introduced. Stacks of alternate layers can be obtained repeating this process many times. McClements et al. deposit a layer of a negatively charged polysaccharide on top of an already adsorbed protein film at low pH, below the pI of the protein [57,58]. The protein film is positively charged below the pI and thus attracts the anionic polysaccharide.

However, the 'layer-by-layer' approach has some disadvantages; in fact, emulsion droplet tends to bridge flocculation and depletion flocculation. Bridging flocculation occurs when the polysaccharide concentration is low since the droplet collisions are faster than the rate of polysaccharide saturation of the protein-coated droplet surfaces. On the other hand, depletion flocculation occurs at higher polysaccharide concentration when a critical value of unadsorbed polysaccharide is exceeded. Due to these disadvantages, mixed emulsions are preferred in practical applications.

Wang, et al. [59] deeply investigated the effect of pH on soy protein isolated-soy hull polysaccharides complex system. The protein adsorbs on the surface of the polysaccharide through electrostatic attraction. The particle size, penetration rate, reorganization rate and dilatational viscoelasticity of the protein-polysaccharide conjugate were reported to increase, while diffusion rate and interfacial pressure decreased with the increasing of pH value from 2 to 4. The amounts of soy hull polysaccharides adsorbed at the oil-water interface reached the maximum at pH 5, while the minimum at pH 6. The zeta-potential was found to be more negative in the pH range between 6 and 8, the authors suggested competitive adsorption between the protein and the polysaccharide. On the other hand, at pH 9 the negative charge on the surface was found to decrease. Their results can be related to the complex behavior of the pI respect to the pH.

Owens et al. [60**] investigated the menhaden oil-in-water emulsion stability containing whey protein isolate and xanthan-locust bean gum as a function of the pH. It is not highlighted in their discussion, but the experimental procedure is a layer-by-layer approach. The emulsions had large particle sizes, viscosity, droplet aggregation, and creaming index, resulting in poor physical stability at pH 3 and 5, while at pH 7, the protein-polysaccharide emulsion was stable due to electrostatic repulsions. The whey protein isolate and xanthan-locust bean gum mixtures were capable to form stable emulsions at pH 3 and pH 7. Bringing flocculation appeared below and around the protein isoelectric point. At pH 7, emulsions containing xanthan-locust bean gum mixtures as secondary emulsifiers were more stable than emulsions with xanthan or locust bean gum alone. The enhanced viscosity that resulted from the polysaccharide mixture interaction may have increased the oxidative stability of the protein-coated droplets at pH 7 by further increasing the electrostatic repulsion between the droplets.

### 3.3. Mixed emulsions

Wang et al. [61*] made a systematic comparison between the hetero-aggregated emulsion, the layer-by-layer and directly mixing techniques. The authors reported that the viscosity of the emulsions stabilized with whey protein isolate and flaxseed gum was in the order of hetero-aggregated emulsion > layer-by-layer emulsion > directly mixing emulsion. Furthermore, they confirmed that emulsion stability is higher for hetero-aggregated emulsion respect to those of layer-by-layer and directly mixing emulsions. They reported that the three-dimensional network structure of hetero-aggregate, observed by cryo-TEM, was the main reason for maintaining high stability and shear rheological properties. However, the three-dimensional structures were sensible to shear flow, while the structures prepared by layer-by-layer and directly mixing, with higher elasticity, were affected in much less extent to the shear flow.

Zhao, et al. [62] investigated calcium sulfate-induced soy protein isolate gels in combination with 0.1, 0.3 and 0.5 % w/v konjac gum, gellan gum, and curdlan gum. In their experimental procedure, a freshly prepared 10% w/v $CaSO_4$ solution was added to the protein-polysaccharide dispersions and stirred. The formation of the gels was obtained by the quenching procedure. The protein-polysaccharide gels revealed higher values of elastic, G′, and viscous, G″, moduli, respect to the protein gel and consequently a better resistance to fracture stress. The network structures of gels were strengthened by the incorporation of the polysaccharides, in particular, the authors reported that curdlan gum exhibited the most significant effects followed by gellan gum and konjac gum. However, the pH effect respect for the different polysaccharides was not considered. The authors also reported a lower gelation temperature of the protein-polysaccharide gels to respect to the protein gel, explaining such evidence with the reinforced protein-protein interactions.

### 3.4. Stabilizing Pickering emulsions

A Pickering emulsion is an emulsion that is stabilized by solid particles that adsorb onto the interface between the two phases [63]. There is a vast amount of research on Pickering emulsions [64–68]. However, it is still a challenge to produce edible particles with the appropriate size (~10–60 nm) and surface chemistry. One of the main challenges is to find the proper stabilizer. Pickering emulsions can be stabilized by polysaccharide-based particles [64,65,69,70] that can provide strong steric stabilization, but lack as emulsifiers. Protein-based particles have been also adopted to stabilize the Pickering emulsions [71–74], even though a structural dissociation can be noticed when adsorbed at the interface. The protein-polysaccharide conjunction can provide both good emulsification performance and strong steric stabilization as demonstrated by recent works [55*,75–79].

Doost et al. [80] adopted soluble coacervates particles made of whey protein isolate and the soluble fraction of almond gum with ratio 3:2 total concentration 0.4 %w/v and pH 4.5 to induce Pickering stabilization of thymol. Static light scattering and microscopy investigation proved the formation of thymol oil-in-water emulsions by adsorption of nano-complexes on the surface of the droplets. The flocculation observed by using the protein at pH 4.5 as a stabilizer was not observed by using the protein-polysaccharide combination at the same pH.

Yang et al. [81] reported a Pickering emulsion stabilization through Okara protein-polysaccharide nanoparticles. The authors used Okara as row material, which is a byproduct of soybean products [82]. Okara contains several insoluble polysaccharides, while the protein composition is comparable to that of soy protein isolate [82,83]. The nanoparticles were fabricated by the ultra-sonication from insoluble soy polysaccharides. The nanoparticles were stable over the pH range 2.0–12.0. The particle concentration of 0.25 wt% was already effective on the stabilization, higher concentrations led to a progressive strengthening of their gel network, as well as a progressive decrease in their droplet size. However, increasing pH the rheological results reported by the authors show a clearly decrease of both elastic and viscous moduli, along with an increase of the particle size consequently changing the volume fraction affecting mechanical properties. All the stabilized gels exhibited excellent stability against prolonged storage and heating, as well as the unique reversibility of freeze-thawing-destabilization/re-emulsification.

## 4. Enzymatic crosslinking

Crosslinking is defined as "the process of forming tridimensional networks", where polymer chains may be linked by covalent or noncovalent bonds. Crosslinking reduces the mobility of the polymer structure and usually enhances its mechanical and barrier properties [84], reducing both its water solubility [85,86] and swelling [87]. Crosslinking reactions are commonly applied to proteins than to polysaccharides since proteins have more functional groups [87]. There are several ways to obtain crosslinking for more details see Azeredo and Waldron [88*], here we reported a recent work on enzymatic crosslinking between protein and polysaccharide since the enzymatic crosslinking is specific and requires mild reaction conditions. Chen et al. [89*] adopted soy protein isolate and sugar beet pectin to fabricate the so-called double network gels (mixed emulsion approach) via thermal treatment and laccase-catalysis. The obtained network is leading to a gel with higher holding water capacity respect to the polysaccharide gel alone. The interpenetrating networks gradually formed with the increase of the protein concentration as seen by qualitative rheological measurements; the gel is more elastic at 8% protein and 2% laccase. The role of laccase was associated to catalyze oxygen to oxidize ferulic acid (in the polysaccharide) and tyrosine (in the protein) [90,91]. Chen et al. demonstrated successfully that enzymatic crosslinking could be adopted to stabilize emulsions. However, the encapsulation and release of active compounds should be evaluated, even though the gel hardness could be adjusted by changing protein or enzyme concentration.

## 5. Conclusions and future perspectives

Even though the food industry still largely uses proteins as emulsion stabilizers and food-grade nanoparticles, this review has examined recent progress on protein-polysaccharide combination to achieve advanced emulsification performance and strong steric and in some cases electrostatic stabilization properties. The complex combinations allow designing an amphiphilic conjugate strongly anchored to the oil–water interface via the protein's hydrophobic regions, while the non-adsorbing polysaccharide region to provide enhanced steric stabilization. However, experimental conditions such as pH, ionic strength, concentration, heat and mechanical treatments should be carefully settled respect to the isoelectric point of the protein and the pKa of the polysaccharide for both conjugates obtained via physical or chemical bonding. It is relevant to avoid Amadori compounds in the case of Maillard conjugates between protein and polysaccharides. While it is relevant to set the appropriate pH for the electrostatic interaction between the protein and the polysaccharide.

There two interesting perspectives related to the protein-polysaccharide topic: i) nutrients encapsulation and controlled delivery; ii) a combination of multiple techniques to obtained more stable and efficient complex aggregates.

There is a significant interest in the food industries to protect bioactive molecules against release in the stomach or intestine [92]. At the same time, the formation and structure of protein-polysaccharide complexes has been widely investigated as well as their functional properties. Several studies are focusing on the encapsulation and delivery of nutrients by using protein-polysaccharide particles due to their enhanced mechanical properties and stability [93–95].

The recent studies working on the coacervates stabilization through Maillard reaction [55*] and protein-polysaccharide gel stabilization thought enzymatic crosslinking [89*] are interesting approaches. In particular, these stabilized structures could be adopted as stable encapsulating agents for nutrients or other components such as colorants, giving at the same time a pleasant texture during consumption. However, developments are needed. The Maillard reaction conditions to stabilize coacervates should be improved to avoid glycate hydrolysis and Amadori compounds like suggested by Li and Etzel [31**]. The elastic properties of the enzymatic cross-linked network should also be optimized for the encapsulation and the subsequent delivery of active compounds.

**References and recommended reading*,****

[1]     Day L, Golding M. Food Structure, Rheology, and Texture. Encyclopedia of Food Chemistry, Elsevier; 2016, p. 125–9. https://doi.org/10.1016/b978-0-08-100596-5.03412-0.
[2]     Borwankar RP. Food texture and rheology: A tutorial review. Journal of Food Engineering 1992;16:1–16. https://doi.org/10.1016/0260-8774(92)90016-Y.
[3]     Bourne MC. Food Texture and Viscosity: 4 Principles of Objective Texture Measurements. 2002.
[4]     Pascua Y, Koç H, Foegeding EA. Food structure: Roles of mechanical properties and oral processing in determining sensory texture of soft materials. Current Opinion in Colloid and Interface Science 2013;18:324–33. https://doi.org/10.1016/j.cocis.2013.03.009.
[5*]    Ettelaie R, Zengin A, Lishchuk S V. Novel food grade dispersants: Review of recent progress. Current Opinion in Colloid and Interface Science 2017;28:46–55. https://doi.org/10.1016/j.cocis.2017.03.004. The authors discussed novel food-grade dispersants that molecular adsorbed at the interface including protein-polysaccharide mixed layers, Maillard conjugates, hydrophobic modification of starch and other polysaccharides and the use of polypeptides obtained from modest hydrolysis of various proteins.
[6*]    Dickinson E. Colloids in Food: Ingredients, Structure, and Stability. Annual Review of Food Science and Technology 2015;6:211–33. https://doi.org/10.1146/annurev-food-022814-015651. Dickinson is providing an interesting overview of new food ingredients to be adopted as emulsifiers, stabilizing and foaming agents, including protein-polysaccharide conjugates.
[7]     Daget N, Joerg M, Bourne M. Creamy Perception I: in Model Dessert Creams. Journal of Texture Studies 1987;18:367–88. https://doi.org/10.1111/j.1745-4603.1987.tb00913.x.
[8]     Akhtar M, Stenzel J, Murray BS, Dickinson E. Factors affecting the perception of creaminess of oil-in-water emulsions. Food Hydrocolloids, vol. 19, 2005, p. 521–6.

https://doi.org/10.1016/j.foodhyd.2004.10.017.

[9] Chen J, Rosenthal AJ. Modifying food texture. Volume 1, Novel ingredients and processing techniques. n.d.

[10] Schmidt KA, Smith DE. Rheological Properties of Gum and Milk Protein Interactions. Journal of Dairy Science 1992;75:36–42. https://doi.org/10.3168/jds.S0022-0302(92)77735-2.

[11] Schorsch C, Jones MG, Norton IT. Thermodynamic incompatibility and microstructure of milk protein/locust bean gum/sucrose systems. Food Hydrocolloids 1999;13:89–99. https://doi.org/10.1016/S0268-005X(98)00074-5.

[12**] Ghosh KA, Bandyopadhyay P. Polysaccharide-Protein Interactions and Their Relevance in Food Colloids. The Complex World of Polysaccharides, InTech; 2012. https://doi.org/10.5772/50561. Ghosh and Bandyopadhyay discussed the interaction between protein and polysaccharide respect to pH of the system, pI of the protein and pKa of the polysaccharide.

[13] Dickinson E. Hydrocolloids at interfaces and the influence on the properties of dispersed systems. Food Hydrocolloids 2003;17:25–39. https://doi.org/10.1016/S0268-005X(01)00120-5.

[14**] Dickinson E. Interfacial structure and stability of food emulsions as affected by protein-polysaccharide interactions. Soft Matter 2008;4:932–42. https://doi.org/10.1039/b718319d. Here Dickinson is providing a wonderful physico-chemical discussion on associative physical interactions leading to electrostatic protein-polysaccharide complexes and covalent Maillard-type conjugates. He is also providing some references on modelling and simulation approaches.

[15**] Bos MA, Van Vliet T. Interfacial rheological properties of adsorbed protein layers and surfactants: A review. Advances in Colloid and Interface Science 2001;91:437–71. https://doi.org/10.1016/S0001-8686(00)00077-4. The authors investigated the relationship between the interfacial rheological properties of adsorbed protein layers and the different stability and instability mechanisms in protein-stabilized and mixed protein–surfactant-stabilized emulsions and foams elucidating at the same time molecular conformation and structure of individual proteins in mixed films.

[16] Sánchez CC, Patino JMR. Interfacial, foaming and emulsifying characteristics of sodium caseinate as influenced by protein concentration in solution. Food Hydrocolloids, vol. 19, 2005, p. 407–16. https://doi.org/10.1016/j.foodhyd.2004.10.007.

[17] Krägel J, Derkatch SR, Miller R. Interfacial shear rheology of protein-surfactant layers. Advances in Colloid and Interface Science 2008;144:38–53. https://doi.org/10.1016/j.cis.2008.08.010.

[18] Koupantsis T, Kiosseoglou V. Whey protein-carboxymethylcellulose interaction in solution and in oil-in-water emulsion systems. Effect on emulsion stability. Food Hydrocolloids 2009;23:1156–63. https://doi.org/10.1016/j.foodhyd.2008.09.004.

[19] Mackie AR. Structure of adsorbed layers of mixtures of proteins and surfactants. Current Opinion in Colloid and Interface Science 2004;9:357–61. https://doi.org/10.1016/j.cocis.2004.08.001.

[20*] de Oliveira FC, Coimbra JS dos R, de Oliveira EB, Zuñiga ADG, Rojas EEG. Food Protein-polysaccharide Conjugates Obtained via the Maillard Reaction: A Review. Critical Reviews in Food Science and Nutrition 2016;56:1108–25. https://doi.org/10.1080/10408398.2012.755669. This critical review is highlighting the properties and the issues related to the Maillard reaction between protein and polysaccharide, discussing the degradation of the Amadori compound formed during the reaction.

[21] Dickinson E. Flocculation of protein-stabilized oil-in-water emulsions. Colloids and Surfaces B: Biointerfaces 2010;81:130–40. https://doi.org/10.1016/j.colsurfb.2010.06.033.

[22] Vaclavik VA, Christian EW. Proteins in Food: An Introduction. Essentials of Food Science, New York, NY: Springer New York; n.d., p. 145–59. https://doi.org/10.1007/978-0-387-69940-0_8.

[23] Belitz HD, Grosch W, Schieberle P. Food chemistry. Springer Berlin Heidelberg; 2009. https://doi.org/10.1007/978-3-540-69934-7.

[24] Pelegrine DHG, Gasparetto CA. Whey proteins solubility as function of temperature and pH. LWT - Food Science and Technology 2005;38:77–80. https://doi.org/10.1016/j.lwt.2004.03.013.

[25]   Sriprablom J, Luangpituksa P, Wongkongkatep J, Pongtharangkul T, Suphantharika M. Influence of pH and ionic strength on the physical and rheological properties and stability of whey protein stabilized o/w emulsions containing xanthan gum. Journal of Food Engineering 2019;242:141–52. https://doi.org/10.1016/j.jfoodeng.2018.08.031.

[26]   Kato A. Maillard-type protein-polysaccharide conjugates. Developments in Food Science 2000;41:385–95. https://doi.org/10.1016/S0167-4501(00)80017-5.

[27]   Akhtar M, Dickinson E. Whey protein-maltodextrin conjugates as emulsifying agents: An alternative to gum arabic. Food Hydrocolloids 2007;21:607–16. https://doi.org/10.1016/j.foodhyd.2005.07.014.

[28]   KATO A. Industrial Applications of Maillard-Type Protein-Polysaccharide Conjugates. Food Science and Technology Research 2002;8:193–9. https://doi.org/10.3136/fstr.8.193.

[29]   Oliver CM, Melton LD, Stanley RA. Creating Proteins with Novel Functionality via the Maillard Reaction: A Review. Critical Reviews in Food Science and Nutrition 2006;46:337–50. https://doi.org/10.1080/10408690590957250.

[30]   Liu J, Ru Q, Ding Y. Glycation a promising method for food protein modification: Physicochemical properties and structure, a review. Food Research International 2012;49:170–83. https://doi.org/10.1016/j.foodres.2012.07.034.

[31**] Li N, Etzel MR. Hydrolysis of Whey Protein-Dextran Glycates Made Using the Maillard Reaction. Foods 2019;8:686. https://doi.org/10.3390/foods8120686. In this brilliant article, the authors are discussing on how to avoid glycate formation and hydrolysis during the Maillard reaction, suggesting a food-grade reducing agent after the formation of the glycate.

[32]   Cordes EH, Jencks WP. **The Mechanism of Hydrolysis of Schiff Bases Derived from Aliphatic Amines**. Journal of the American Chemical Society 1963;85:2843–8. https://doi.org/10.1021/ja00901a037.

[33]   Kutzli I, Griener D, Gibis M, Schmid C, Dawid C, Baier SK, et al. Influence of Maillard reaction conditions on the formation and solubility of pea protein isolate-maltodextrin conjugates in electrospun fibers. Food Hydrocolloids 2019:105535. https://doi.org/10.1016/J.FOODHYD.2019.105535.

[34*]  Zha F, Dong S, Rao J, Chen B. Pea protein isolate-gum Arabic Maillard conjugates improves physical and oxidative stability of oil-in-water emulsions. Food Chemistry 2019;285:130–8. https://doi.org/10.1016/j.foodchem.2019.01.151. The authors were able to enhance protein solubility by controlling the Maillard reaction of pea protein isolate with gum Arabic, in particular, focusing on the incubation time. See the text for further details.

[35*]  Zhong L, Ma N, Wu Y, Zhao L, Ma G, Pei F, et al. Characterization and functional evaluation of oat protein isolate-Pleurotus ostreatus β-glucan conjugates formed via Maillard reaction. Food Hydrocolloids 2019;87:459–69. https://doi.org/10.1016/j.foodhyd.2018.08.034. The authors proved that the Maillard reaction could enhance the utilization of oat protein isolate in the food industry under controlled dry-heating conditions. See the text for further details.

[36]   Lertittikul W, Benjakul S, Tanaka M. Characteristics and antioxidative activity of Maillard reaction products from a porcine plasma protein-glucose model system as influenced by pH. Food Chemistry 2007;100:669–77. https://doi.org/10.1016/j.foodchem.2005.09.085.

[37]   Dickinson E. Hydrocolloids as emulsifiers and emulsion stabilizers. Food Hydrocolloids 2009;23:1473–82. https://doi.org/10.1016/j.foodhyd.2008.08.005.

[38]   Guo X, Xiong YL. Characteristics and functional properties of buckwheat protein-sugar Schiff base complexes. LWT - Food Science and Technology 2013;51:397–404. https://doi.org/10.1016/j.lwt.2012.12.003.

[39]   Qi J, Liao J, Yin S, Zhu J, Yang X. Formation of acid-precipitated soy protein-dextran conjugates by Maillard reaction in liquid systems. International Journal of Food Science & Technology 2010;45:2573–80. https://doi.org/10.1111/j.1365-2621.2010.02433.x.

[40]     Bengoechea C, Peinado I, McClements DJ. Formation of protein nanoparticles by controlled heat treatment of lactoferrin: Factors affecting particle characteristics. Food Hydrocolloids 2011;25:1354–60. https://doi.org/10.1016/j.foodhyd.2010.12.014.
[41]     Stone AK, Nickerson MT. Formation and functionality of whey protein isolate-(kappa-, iota-, and lambda-type) carrageenan electrostatic complexes. Food Hydrocolloids 2012;27:271–7. https://doi.org/10.1016/j.foodhyd.2011.08.006.
[42]     McClements DJ. Non-covalent interactions between proteins and polysaccharides. Biotechnology Advances 2006;24:621–5. https://doi.org/10.1016/j.biotechadv.2006.07.003.
[43**]   Moschakis T, Biliaderis CG. Biopolymer-based coacervates: Structures, functionality and applications in food products. Current Opinion in Colloid and Interface Science 2017;28:96–109. https://doi.org/10.1016/j.cocis.2017.03.006. Moschakis and Biliaderis proposed an extended discussion on coacervates and complex coacervates focusing on their structure as well as on their ability to encapsulate and protect sensitive ingredients from degradation.
[44]     De Kruif CG, Weinbreck F, De Vries R. Complex coacervation of proteins and anionic polysaccharides. Current Opinion in Colloid and Interface Science 2004;9:340–9. https://doi.org/10.1016/j.cocis.2004.09.006.
[45]     Nigen M, Croguennec T, Renard D, Bouhallab S. Temperature Affects the Supramolecular Structures Resulting from α-Lactalbumin−Lysozyme Interaction. Biochemistry 2007;46:1248–55. https://doi.org/10.1021/bi062129c.
[46]     Clarke AW, Arnspang EC, Mithieux SM, Korkmaz E, Braet F, Weiss AS. Tropoelastin Massively Associates during Coacervation To Form Quantized Protein Spheres †. Biochemistry 2006;45:9989–96. https://doi.org/10.1021/bi0610092.
[47]     Croguennec T, Tavares GM, Bouhallab S. Heteroprotein complex coacervation: A generic process. Advances in Colloid and Interface Science 2017;239:115–26. https://doi.org/10.1016/j.cis.2016.06.009.
[48]     Kim S, Huang J, Lee Y, Dutta S, Young Yoo H, Mee Jung Y, et al. Complexation and coacervation of like-charged polyelectrolytes inspired by mussels. Proceedings of the National Academy of Sciences of the United States of America 2016;113:E847–53. https://doi.org/10.1073/pnas.1521521113.
[49]     Schmitt C, Turgeon SL. Protein/polysaccharide complexes and coacervates in food systems. Advances in Colloid and Interface Science 2011;167:63–70. https://doi.org/10.1016/j.cis.2010.10.001.
[50]     Cooper CL, Dubin PL, Kayitmazer AB, Turksen S. Polyelectrolyte-protein complexes. Current Opinion in Colloid and Interface Science 2005;10:52–78. https://doi.org/10.1016/j.cocis.2005.05.007.
[51]     Kayitmazer AB, Seeman D, Minsky BB, Dubin PL, Xu Y. Protein-polyelectrolyte interactions. Soft Matter 2013;9:2553–83. https://doi.org/10.1039/c2sm27002a.
[52]     Lan Y, Ohm JB, Chen B, Rao J. Phase behavior and complex coacervation of concentrated pea protein isolate-beet pectin solution. Food Chemistry 2020;307. https://doi.org/10.1016/j.foodchem.2019.125536.
[53]     Chang PG, Gupta R, Timilsena YP. Rheological and Microstructural Characteristics of Canola Protein Isolate−Chitosan Complex Coacervates. Journal of Food Science 2019;84:1104–12. https://doi.org/10.1111/1750-3841.14599.
[54*]    Wu C, Chen Q, Li X, Su J, He S, Liu J, et al. Formation and characterisation of food protein–polysaccharide thermal complex particles: effects of pH, temperature and polysaccharide type. International Journal of Food Science & Technology 2019:ijfs.14416. https://doi.org/10.1111/ijfs.14416. The authors proposed a phase diagram with respect to the pH and polysaccharide type, in particular, they were looking at soybean protein isolate interaction with chitosan and carboxymethyl cellulose. They also considered temperature in their study. See the text for further details.
[55*]    Huang G-Q, Wang H-O, Wang F-W, Du Y-L, Xiao J-X. Maillard reaction in protein – polysaccharide

coacervated microcapsules and its effects on microcapsule properties. International Journal of Biological Macromolecules 2019. https://doi.org/10.1016/j.ijbiomac.2019.11.087. The authors proposed to use a Maillard reaction to stabilize coacervates, they demonstrated that the reaction did not destroy the core-shell structure of the microcapsules, on the other hand, improved the microencapsulation performance. See the text for further details.

[56]  Decher G. Fuzzy nanoassemblies: Toward layered polymeric multicomposites. Science 1997;277:1232–7. https://doi.org/10.1126/science.277.5330.1232.

[57]  Guzey D, McClements DJ. Formation, stability and properties of multilayer emulsions for application in the food industry. Advances in Colloid and Interface Science 2006;128–130:227–48. https://doi.org/10.1016/j.cis.2006.11.021.

[58]  Guzey D, McClements DJ. Impact of Electrostatic Interactions on Formation and Stability of Emulsions Containing Oil Droplets Coated by β-Lactoglobulin−Pectin Complexes. Journal of Agricultural and Food Chemistry 2007;55:475–85. https://doi.org/10.1021/jf062342f.

[59]  Wang S, Yang J, Shao G, Qu D, Zhao H, Zhu L, et al. Dilatational rheological and nuclear magnetic resonance characterization of oil-water interface: Impact of pH on interaction of soy protein isolated and soy hull polysaccharides. Food Hydrocolloids 2020;99. https://doi.org/10.1016/j.foodhyd.2019.105366.

[60**]  Owens C, Griffin K, Khouryieh H, Williams K. Creaming and oxidative stability of fish oil-in-water emulsions stabilized by whey protein-xanthan-locust bean complexes: Impact of pH. Food Chemistry 2018;239:314–22. https://doi.org/10.1016/j.foodchem.2017.06.096. The authors investigated a combination of polysaccharide (xanthan and locust bean gum) to interact with whey protein isolate as a function of pH. See the text for further details.

[61*]  Wang X, Li X, Xu D, Zhu Y, Cao Y, Wang J, et al. Comparision of heteroaggregation, layer-by-layer and directly mixing techniques on the physical properties and in vitro digestion of emulsions. Food Hydrocolloids 2019;95:228–37. https://doi.org/10.1016/j.foodhyd.2019.04.034. The authors made an interesting comparison between heteroaggregation, layer-by-layer and directly mixing techniques on the preparation of the whey protein isolate and flaxseed gum complex by analyzing mainly the rheological behavior of the emulsions. See the text for further details.

[62]  Zhao H, Chen J, Hemar Y, Cui B. Improvement of the rheological and textural properties of calcium sulfate-induced soy protein isolate gels by the incorporation of different polysaccharides. Food Chemistry 2020;310. https://doi.org/10.1016/j.foodchem.2019.125983.

[63]  Pickering SU. CXCVI. - Emulsions. Journal of the Chemical Society, Transactions 1907;91:2001–21. https://doi.org/10.1039/CT9079102001.

[64]  Rayner M, Timgren A, Sjöö M, Dejmek P. Quinoa starch granules: a candidate for stabilising food-grade Pickering emulsions. Journal of the Science of Food and Agriculture 2012;92:1841–7. https://doi.org/10.1002/jsfa.5610.

[65]  Tan Y, Xu K, Liu C, Li Y, Lu C, Wang P. Fabrication of starch-based nanospheres to stabilize pickering emulsion. Carbohydrate Polymers 2012;88:1358–63. https://doi.org/10.1016/j.carbpol.2012.02.018.

[66]  Kargar M, Fayazmanesh K, Alavi M, Spyropoulos F, Norton IT. Investigation into the potential ability of Pickering emulsions (food-grade particles) to enhance the oxidative stability of oil-in-water emulsions. Journal of Colloid and Interface Science 2012;366:209–15. https://doi.org/10.1016/j.jcis.2011.09.073.

[67]  de Castro Santana R, Kawazoe Sato AC, Lopes da Cunha R. Emulsions stabilized by heat-treated collagen fibers. Food Hydrocolloids 2012;26:73–81. https://doi.org/10.1016/j.foodhyd.2011.04.006.

[68]  Luo Z, Murray BS, Ross AL, Povey MJW, Morgan MRA, Day AJ. Effects of pH on the ability of flavonoids to act as Pickering emulsion stabilizers. Colloids and Surfaces B: Biointerfaces 2012;92:84–90. https://doi.org/10.1016/j.colsurfb.2011.11.027.

[69]   Capron I, Cathala B. Surfactant-Free High Internal Phase Emulsions Stabilized by Cellulose Nanocrystals. Biomacromolecules 2013;14:291–6. https://doi.org/10.1021/bm301871k.

[70]   Chen QH, Zheng J, Xu YT, Yin SW, Liu F, Tang CH. Surface modification improves fabrication of pickering high internal phase emulsions stabilized by cellulose nanocrystals. Food Hydrocolloids 2018;75:125–30. https://doi.org/10.1016/j.foodhyd.2017.09.005.

[71]   Hu YQ, Yin SW, Zhu JH, Qi JR, Guo J, Wu LY, et al. Fabrication and characterization of novel Pickering emulsions and Pickering high internal emulsions stabilized by gliadin colloidal particles. Food Hydrocolloids 2016;61:300–10. https://doi.org/10.1016/j.foodhyd.2016.05.028.

[72]   Jiao C, Zhang Z, Zhong C, Feng Z. An indoor mmWave joint radar and communication system with active channel perception. IEEE International Conference on Communications, vol. 2018- May, Institute of Electrical and Electronics Engineers Inc.; 2018. https://doi.org/10.1109/ICC.2018.8422132.

[73]   Xu Y-T, Tang C-H, Liu T-X, Liu R. Ovalbumin as an Outstanding Pickering Nanostabilizer for High Internal Phase Emulsions. Journal of Agricultural and Food Chemistry 2018;66:8795–804. https://doi.org/10.1021/acs.jafc.8b02183.

[74]   Xu YT, Liu TX, Tang CH. Novel pickering high internal phase emulsion gels stabilized solely by soy β-conglycinin. Food Hydrocolloids 2019;88:21–30. https://doi.org/10.1016/j.foodhyd.2018.09.031.

[75]   Xu Y, Atrens AD, Stokes JR. Rheology and microstructure of aqueous suspensions of nanocrystalline cellulose rods. Journal of Colloid and Interface Science 2017;496:130–40. https://doi.org/10.1016/j.jcis.2017.02.020.

[76]   Liu F, Zheng J, Huang CH, Tang CH, Ou SY. Pickering high internal phase emulsions stabilized by protein-covered cellulose nanocrystals. Food Hydrocolloids 2018;82:96–105. https://doi.org/10.1016/j.foodhyd.2018.03.047.

[77]   Wijaya W, Van der Meeren P, Wijaya CH, Patel AR. High internal phase emulsions stabilized solely by whey protein isolate-low methoxyl pectin complexes: effect of pH and polymer concentration. Food and Function 2017;8:584–94. https://doi.org/10.1039/c6fo01027j.

[78]   Zhou FZ, Zeng T, Yin SW, Tang CH, Yuan DB, Yang XQ. Development of antioxidant gliadin particle stabilized Pickering high internal phase emulsions (HIPEs) as oral delivery systems and the: In vitro digestion fate. Food and Function 2018;9:959–70. https://doi.org/10.1039/c7fo01400g.

[79]   Zeng T, Wu Z ling, Zhu JY, Yin SW, Tang CH, Wu LY, et al. Development of antioxidant Pickering high internal phase emulsions (HIPEs) stabilized by protein/polysaccharide hybrid particles as potential alternative for PHOs. Food Chemistry 2017;231:122–30. https://doi.org/10.1016/j.foodchem.2017.03.116.

[80]   Sedaghat Doost A, Nikbakht Nasrabadi M, Kassozi V, Dewettinck K, Stevens C V., Van der Meeren P. Pickering stabilization of thymol through green emulsification using soluble fraction of almond gum – Whey protein isolate nano-complexes. Food Hydrocolloids 2019;88:218–27. https://doi.org/10.1016/j.foodhyd.2018.10.009.

[81]   Yang T, Li XT, Tang CH. Novel edible pickering high-internal-phase-emulsion gels efficiently stabilized by unique polysaccharide-protein hybrid nanoparticles from Okara. Food Hydrocolloids 2020;98. https://doi.org/10.1016/j.foodhyd.2019.105285.

[82]   Ullah I, Yin T, Xiong S, Zhang J, Din Z ud, Zhang M. Structural characteristics and physicochemical properties of okara (soybean residue) insoluble dietary fiber modified by high-energy wet media milling. LWT - Food Science and Technology 2017;82:15–22. https://doi.org/10.1016/j.lwt.2017.04.014.

[83]   Stanojevic SP, Barac MB, Pesic MB, Vucelic-Radovic B V. Composition of proteins in okara as a byproduct in hydrothermal processing of soy milk. Journal of Agricultural and Food Chemistry, vol. 60, 2012, p. 9221–8. https://doi.org/10.1021/jf3004459.

[84]   Balaguer MP, Gómez-Estaca J, Gavara R, Hernandez-Munoz P. Functional Properties of Bioplastics

- [ ] Made from Wheat Gliadins Modified with Cinnamaldehyde. Journal of Agricultural and Food Chemistry 2011;59:6689–95. https://doi.org/10.1021/jf200477a.
- [85] De Carvalho RA, Grosso CRF. Characterization of gelatin based films modified with transglutaminase, glyoxal and formaldehyde. Food Hydrocolloids 2004;18:717–26. https://doi.org/10.1016/j.foodhyd.2003.10.005.
- [86] Menzel C, Olsson E, Plivelic TS, Andersson R, Johansson C, Kuktaite R, et al. Molecular structure of citric acid cross-linked starch films. Carbohydrate Polymers 2013;96:270–6. https://doi.org/10.1016/j.carbpol.2013.03.044.
- [87] Boanini E, Rubini K, Panzavolta S, Bigi A. Chemico-physical characterization of gelatin films modified with oxidized alginate. Acta Biomaterialia 2010;6:383–8. https://doi.org/10.1016/j.actbio.2009.06.015.
- [88*] Azeredo HMC, Waldron KW. Crosslinking in polysaccharide and protein films and coatings for food contact - A review. Trends in Food Science and Technology 2016;52:109–22. https://doi.org/10.1016/j.tifs.2016.04.008. This is a great review of the main crosslinking agents/mechanisms to improve the performance and applicability of protein- and polysaccharide-based food contact materials. The authors considered both chemical and physical crosslinking agents and mechanisms.
- [89*] Chen H, Gan J, Ji A, Song S, Yin L. Development of double network gels based on soy protein isolate and sugar beet pectin induced by thermal treatment and laccase catalysis. Food Chemistry 2019;292:188–96. https://doi.org/10.1016/j.foodchem.2019.04.059. The authors adopted soy protein isolate and sugar beet pectin to fabricate double network gels via a two-step gelation method (thermal treatment and laccase-catalysis), reporting mechanical, WHC and microstructural characterization. See the text for further details.
- [90] Qiu S, Yadav MP, Yin L. Characterization and functionalities study of hemicellulose and cellulose components isolated from sorghum bran, bagasse and biomass. Food Chemistry 2017;230:225–33. https://doi.org/10.1016/j.foodchem.2017.03.028.
- [91] Zaidel DNA, Chronakis IS, Meyer AS. Enzyme catalyzed oxidative gelation of sugar beet pectin: Kinetics and rheology. Food Hydrocolloids 2012;28:130–40. https://doi.org/10.1016/j.foodhyd.2011.12.015.
- [92] Singh H, Ye A, Horne D. Structuring food emulsions in the gastrointestinal tract to modify lipid digestion. Progress in Lipid Research 2009;48:92–100. https://doi.org/10.1016/j.plipres.2008.12.001.
- [93] Kleemann C, Schuster R, Rosenecker E, Selmer I, Smirnova I, Kulozik U. In-vitro-digestion and swelling kinetics of whey protein, egg white protein and sodium caseinate aerogels. Food Hydrocolloids 2020;101. https://doi.org/10.1016/j.foodhyd.2019.105534.
- [94] Wei Z, Huang Q. Assembly of Protein-Polysaccharide Complexes for Delivery of Bioactive Ingredients: A Perspective Paper. Journal of Agricultural and Food Chemistry 2019;67:1344–52. https://doi.org/10.1021/acs.jafc.8b06063.
- [95] Semenova MG, Moiseenko D V., Grigorovich N V., Anokhina MS, Antipova AS, Belyakova LE, et al. Protein-Polysaccharide Interactions and Digestion of the Complex Particles. Food Structures, Digestion and Health, Elsevier Inc.; 2014, p. 169–92. https://doi.org/10.1016/B978-0-12-404610-8.00006-2.